# Experimental and DFT investigation of $(Cr,Ti)_3AlC_2$ MAX phases stability.

Patrick A. Burr*[,a], Denis Horlait[b,c], William E. Lee[c]

[a] School of EE&T, University of New South Wales, Kensington, NSW, 2052, Australia.

[b] CNRS/IN2P3 and University of Bordeaux, Centre d'Etudes Nucléaires de Bordeaux-Gradignan, UMR 5797, Chemin du Solarium, 33175 Gradignan, France.

[c] Centre for Nuclear Engineering (CNE) & Department of Materials, Imperial College London, SW7 2AZ, UK

* Corresponding author: p.burr@unsw.edu.au.

## Abstract

Using a synergistic combination of experimental and computational methods, we shed light on the unusual solubility of $(Cr,Ti)_3AlC_2$ MAX phase, showing that it may accommodate Cr only at very low concentrations (<2at%) or at the exact Cr/(Cr+Ti) ratio of 2/3, even when the ratio of reactants is far from this stoichiometry (1/2 ≤ Cr/(Cr+Ti) ≤ 5/6). In both phases Cr exclusively occupies the *4f* sites, bridging carbide layers with the Al layer. Despite this, the peculiar stability of $(Cr_{2/3}Ti_{1/3})_3AlC_2$ is attributed to the formation of strong, spin-polarised Cr-C bonds, which result in volume reduction and marked increase in *c/a* ratio.

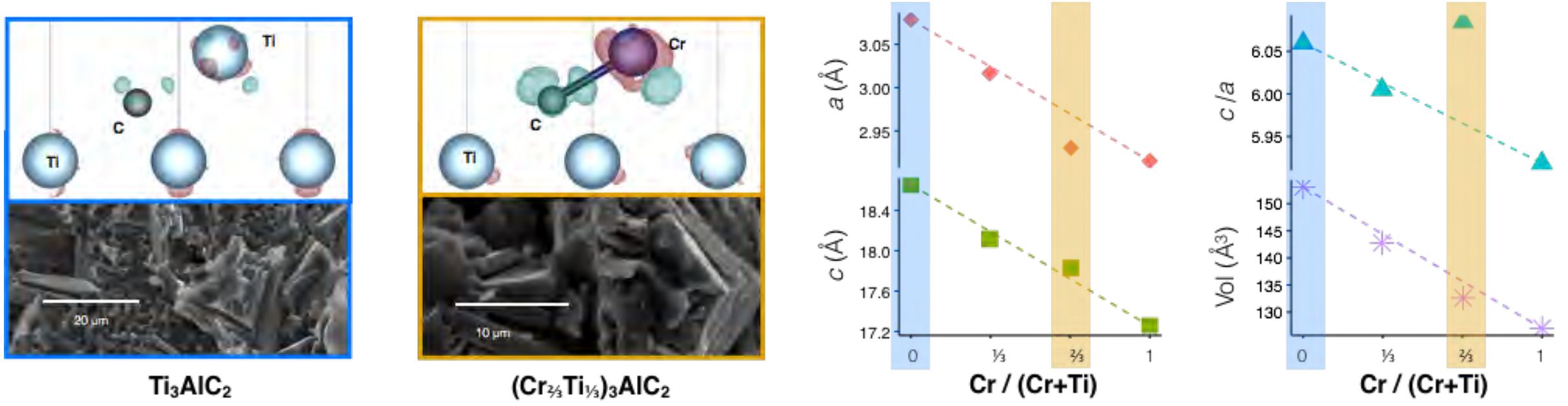

## Keywords

Quaternary MAX phase; Synthesis and characterization; DFT simulations; Atomic ordering.

## Impact statement

Solubility of Cr and Ti in $(Cr,Ti)_3AlC_2$ was investigated using experimental and DFT techniques. It was also determined that $(Cr_{2/3}Ti_{1/3})_3AlC_2$ owe its remarkable stability to the formation strong Cr-C bonds.

## 1. Introduction

$M_{n+1}AX_n$ compounds crystalizing with the $P6_3/mmc$ space group and where M is an early transition metal, A is a group 13-16 element, X is carbon or nitrogen and $n$ is an integer, constitute a material family called MAX phases. The MAX phases are known for having a particular combination of properties owing to the duality of their structure. Indeed, MAX phase structure consists of the stacking of $n$ "ceramic" M-X planes interposed with a "metallic" A layer. As a consequence, MAX phases exhibit a favourable combination of metallic properties (high thermal shock resistance, high thermal and electrical conductivities and good machinability) and ceramic properties (high decomposition / melting temperature and elastic stiffness). [1]

Over 70 ternary MAX phases have been synthesized and a fast-growing number of quaternary MAX phases, [2,3] i.e. containing either two M, two A or two X elements. The interest in quaternary MAX phases is multiple:

- Firstly, they can be used to tailor a certain property to get a value in-between or even better than that of the ternary end-members [4-9]. As an example, Cabioch et al. [9] showed that when substituting 25% of the Al atoms for Ge in $Cr_2AlC$, its thermal expansion becomes isotropic, presumably diminishing thermal stresses in the material.
- Secondly the addition of a fourth element may trigger new properties not observed in the ternary host system. For instance, incorporating Mn in $Cr_2GeC$ [10-12] or $Cr_2GaC$ [4,13-16] was found to render these ternary MAX phases ferromagnetic.
- Thirdly, it allows integrating elements that are not able to form a bulk MAX phase by themselves. In a recent publication we showed that the bismuth containing $Zr_2(Al_{0.42}Bi_{0.58})C$ MAX phase can be synthesized despite no $M_{n+1}BiX_n$ MAX phase being known [17]. Similarly, the recently reported $Zr_2(Sb_{0.7}Al_{0.3})C$ represents the firstly reported antimony-based MAX phase [3].
- Fourthly, it is also interesting on a fundamental basis to help understand the reasons driving the stability (or non-stability) of MAX phases, as this matter is still open to debate [1,18,19].
- Lastly, it allows synthesis of MAX phases with a composition as close as possible to a non-stable ternary one. For example, it can be conservatively said that $Cr_3AlC_2$ cannot be formed as a bulk phase, possibly because $Cr_2AlC$ is more stable [18,20,21]. However, replacing only 1/3 of the chromium atoms for titanium atoms leads to the remarkable MAX phase $(Cr_{2/3}Ti_{1/3})_3AlC_2$ [22,23], which may be stabilized by the ordering of Cr and Ti [24] (Cr atoms exclusively on the 4*f* Wyckoff site, Ti exclusively on the 2*a* sites, as illustrated in Figure 1).

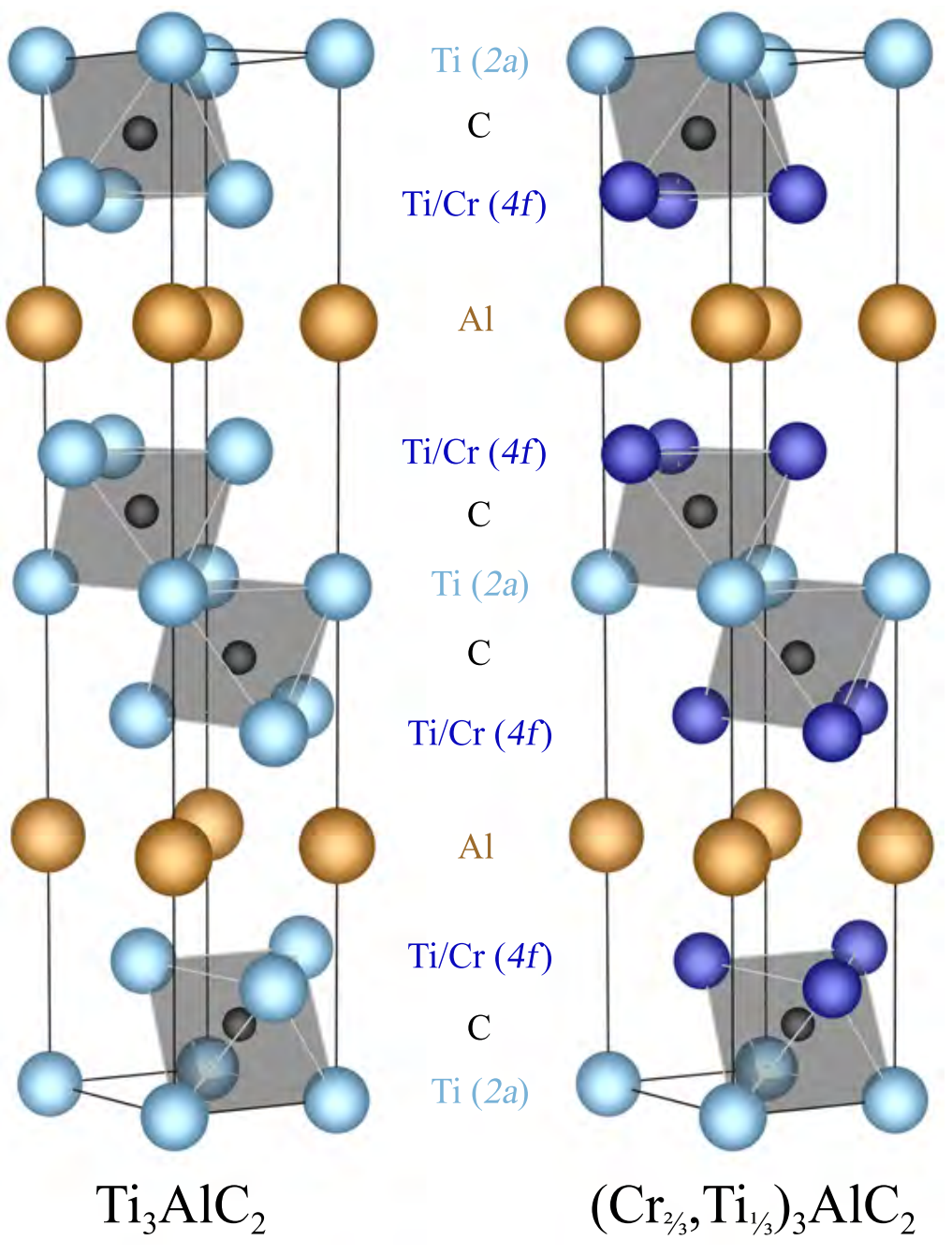

Figure 1 – Crystal structure of $(Cr,Ti)_3AlC_2$ MAX phases.

Continuing on this last example we decided to investigate the stability of quaternary *312*-MAX phases across the compositional range from $Ti_3AlC_2$ to $Cr_3AlC_2$. This was done through a combination of computational methods and experimental syntheses attempts. First the solubility of Cr in $Ti_3AlC_2$ is considered at increasing levels of Cr and the relevant accommodation mechanism is revealed. Then the driving force for ordering in $(Cr_{2/3}Ti_{1/3})_3AlC_2$ is investigated together with its ability to accommodate non-stoichiometry. The source of the peculiar stability of $(Cr_{2/3}Ti_{1/3})_3AlC_2$ is then discussed in detail.

## 2. Methodology

*Experimental details*

Full details of the experimental procedure employed for ceramic synthesis and characterization are found in our previous publication [23]. In brief, $TiH_2$, Cr, Al and graphite powders are employed. All syntheses stages are done under argon atmosphere (except for the introduction at room temperature of the blended powders in the furnace) to limit oxygen contamination. First the elemental powders are mixed by ball-milling (Nanjing University Instrument Plant) according to the ratio Ti+Cr/Al/C = 3/1.1/1. Then the powder mixes are poured in graphite crucibles and placed in a hot-press (FCT Systeme HP W/25/1, Rauenstein, Germany) used as a regular furnace (pressureless mode). The furnace is heated up to 1450°C at a ~20°C.min$^{-1}$ rate and held at the target temperature for 1 hour. The furnace is allowed to cool to room temperature before opening it to ambient atmosphere. Sintering of the reacted powder was then performed by spark plasma sintering (FCT Systeme HP D/25/1 equipment) at 1220°C and 35 MPa. X-ray diffraction (XRD) characterization was performed with a Bruker D2 Phaser SSD160 (Karlsruhe, Germany) and the treatment of results was done with Fullprof Suite program and Xpert High Score Plus software (PANalytical, Almelo, the Netherlands) using the implemented ICDD (International Center for Diffraction Data) database. SEM (scanning electron microscopy) observations were realized using a JEOL JSM-6400 (Tokyo, Japan), operated at 20 keV. The same microscope was used for EDX (energy dispersive X-ray spectrometry) analyses using an INCA ultra-thin polymer window detector (Oxford Instruments, Oxford, U.K.). The samples for SEM/EDX were prepared by embedding pieces of the sintered pellets in a resin then grinding by successive SiC papers down to 5 µm and finally polishing with 3 µm then 1 µm diamond paste. The as-prepared resin blocks are finally gold-coated (~10 nm thick) before the analyses. Further details on the characterization are found in Ref. [23]. When reporting statistical uncertainties of EDX measurements we use the symbol ± to indicate maximum deviation from average and the Greek letter σ to indicate standard deviation.

*Computational methodology*

All DFT simulations were carried out using the Castep code [25,26] employing the PBE exchange-correlation functional [27] and ultra-soft pseudo potentials [28] with a consistent plane-wave energy cut-off of 450 eV. Point defects were investigated in a 3×3×1 supercell (108 atoms) of the conventional unit cell of $M_3AX_2$, in conjunction with the *aneto* finite size correction [29]. Defect clusters were simulated in a 4×4×1 supercell (192 atoms) instead. To test the accuracy of these supercells, selected defects were simulated in both supercells: the difference in energy was consistently below 0.02 eV for point defects, whilst it was as high as 0.10 eV for defect clusters (details provided in section 3). *k*-point spacing was kept as close as possible to 0.03 $Å^{-1}$ throughout, which equates to Monkhors-Pack [30] grids of 3×3×2, 4×4×2 and 10×10×2 for the 192-atom supercell, 108-atom supercell and the unit cell respectively. No symmetry operations were enforced when calculating point defects and all simulations were spin-polarized. Since these systems are metallic, density mixing and cold smearing of bands (broadening width = 0.1 eV) were used. Electronic densities of states were analyzed with the aid of the Optados program [31] with the adaptive broadening scheme [32].

## 3. Dilute Cr incorporation in $Ti_3AlC_2$.

The incorporation of Cr in the $Ti_3AlC_2$ MAX phase [1,33-35] was tried with 5 and 33% substitution of Cr for Ti (target compositions $(Cr_{0.05}Ti_{0.95})_3AlC_2$ and $(Cr_{1/3}Ti_{2/3})_3AlC_2$, respectively). The samples are labeled as $Cr_5Ti_{95}$ and $Cr_{1/3}Ti_{2/3}$, respectively. According to XRD and after matching with the ICDD database, the $Cr_5Ti_{95}$ sample contained mainly a *312* MAX phase in the α polymorph structure (~85 vol%) accompanied with TiC (~15 vol%) [23], the relative volume fraction being determined by Rietveld refinement. The unit cell parameters associated with the obtained MAX phase (a = 3.0744(9) Å; c = 18.567(7) Å) are in the range of literature values for $Ti_3AlC_2$ (3.065 ≤ a ≤ 3.0786 Å and 18.487 ≤ c ≤ 18.73 Å) [34-38]. To investigate if significant amounts of Cr entered the α-$Ti_3AlC_2$ structure, EDX measurements were performed on the sintered $Cr_5Ti_{95}$ sample. Global analyses (~100x100 μm square area) confirmed the expected Ti/Cr/Al ratios, while C was not measured (because of the low atomic mass of C which precludes precise quantification by EDX and because of the probable contamination by SiC and diamond during polishing). Point measurements were then done on individual grains; the corresponding results are presented in Figure 2 (a). The micrograph taken in back-scattered electron (BSE) mode shows three different phases:

- The darker phase (i.e. the phase containing the lightest elements by average atomic number and thereby likely to contain significant amount of C) was found to contain primarily Ti, possibly with some Al and/or Cr, however, the Al and Cr signal may be due to signal contamination by the surrounding grains as the grain size of this dark phase is of the same order as the EDX analysis interaction volume (~1 $μm^3$). As this phase is lighter than the other in the sample and as XRD determined that $Cr_5Ti_{95}$ contains ~15 vol% of TiC, it is safe to assume this first phase to be TiC.
- The main gray phase is expected to be the *312* MAX phase due to the XRD results and the lamellar shape of the grains. The seven point analyses done on different grains of this phase in the area presented in Figure 2 (a) have sensibly the same Al/(Al+Cr+Ti) ratio of 24 at% (with a maximum deviation of the results from the average value of ± 1 at%), which agrees with the expected ratio for a *312* MAX phase. The Cr/(Cr+Ti) ratio is of 1.7 ± 0.3 at%. When combined with the 15-point analysis performed in other areas of the SEM sample, the average value for Cr/(Cr+Ti) becomes 2 ± 1 at%. According to EDX analyses and in conformity with the unit cell parameters of the MAX phase found to be equivalent to that usually reported for $Ti_3AlC_2$ [34-38] the Cr incorporation in $Ti_3AlC_2$ appears to be limited to a maximum of 2 at% of replacement of Ti by Cr, thereby forming $(Cr_{0.02}Ti_{0.98})_3AlC_2$.
- The presence of a brighter Cr-rich third and last phase corroborates the above findings. This phase is observed by SEM-BSE imaging and associated EDX but not by XRD. Excluding C, which is not detected, the composition of these white grains was found to be close to that of $AlCr_2$ with some Cr deficiency and small amounts of Ti (5–14 at%). As the phase has very bright contrast compared to the other phases in BSE imaging and is thus "heavier", it is very likely that the phase does not contain much carbon and it is thus labeled as a metallic $AlCr_{1.8}Ti_{0.2}$. XRD was not able to detect this phase either because it constituted too small a volume fraction (< ~5 vol%) or because it is amorphous.

The synthesis results for *$Cr_{1/3}Ti_{2/3}$* are presented in Ref. [23]. Knowing at this stage the synthesis results of the previous compound, it was anticipated Cr and Ti would split into two separate MAX phases (e.g. $Cr_2AlC$ + $Ti_3AlC_2$ or $(Cr_{2/3}Ti_{1/3})_3AlC_2$ + $Ti_3AlC_2$), however not a single MAX phase was found by XRD analysis and the *$Cr_{1/3}Ti_{2/3}$* only consists of ~65 vol% TiC + ~35 vol% Al-Cr based alloys. This shows that when increased to a certain level, Cr not only does not enter the $Ti_3AlC_2$ structure but also precludes its formation, leading to this TiC + $Al_xCr_y$ composite.

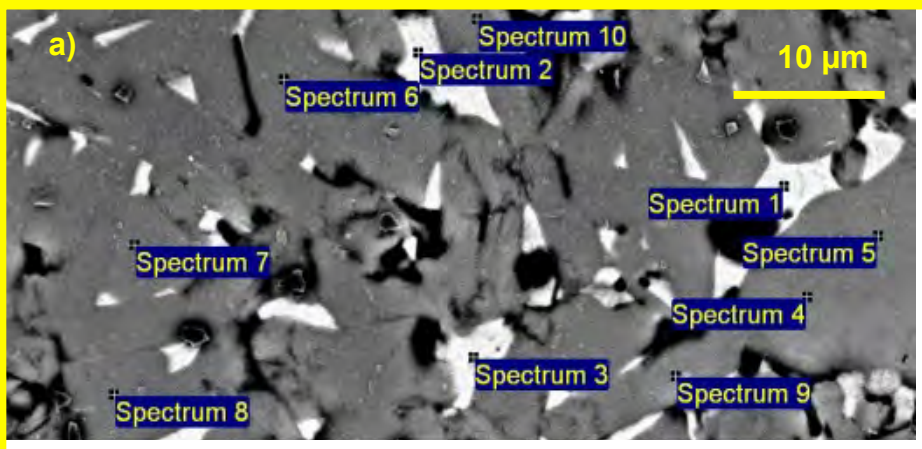

| Spectrum # | **Al**/(Al+Ti+Cr) (%) | **Ti**/(Al+Ti+Cr) (%) | **Cr**/(Al+Ti+Cr) (%) | Determined composition |
|---|---|---|---|---|
| 1 | 34,6 | 6,4 | 59,0 | $AlCr_{1.8}Ti_{0.2}$ |
| 2 | 34,6 | 6,9 | 58,5 | |
| 3 | 33,8 | 12,6 | 53,6 | |
| 4 | 23,5 | 75,2 | 1,3 | $(Cr_{0.02}Ti_{0.98})_3AlC_2$ |
| 5 | 23,3 | 75,5 | 1,2 | |
| 6 | 24,2 | 74,7 | 1,2 | |
| 7 | 24,1 | 74,5 | 1,4 | |
| 8 | 23,4 | 75,5 | 1,1 | |
| 9 | 23,5 | 75,0 | 1,4 | |
| 10 | 25,0 | 73,7 | 1,3 | |

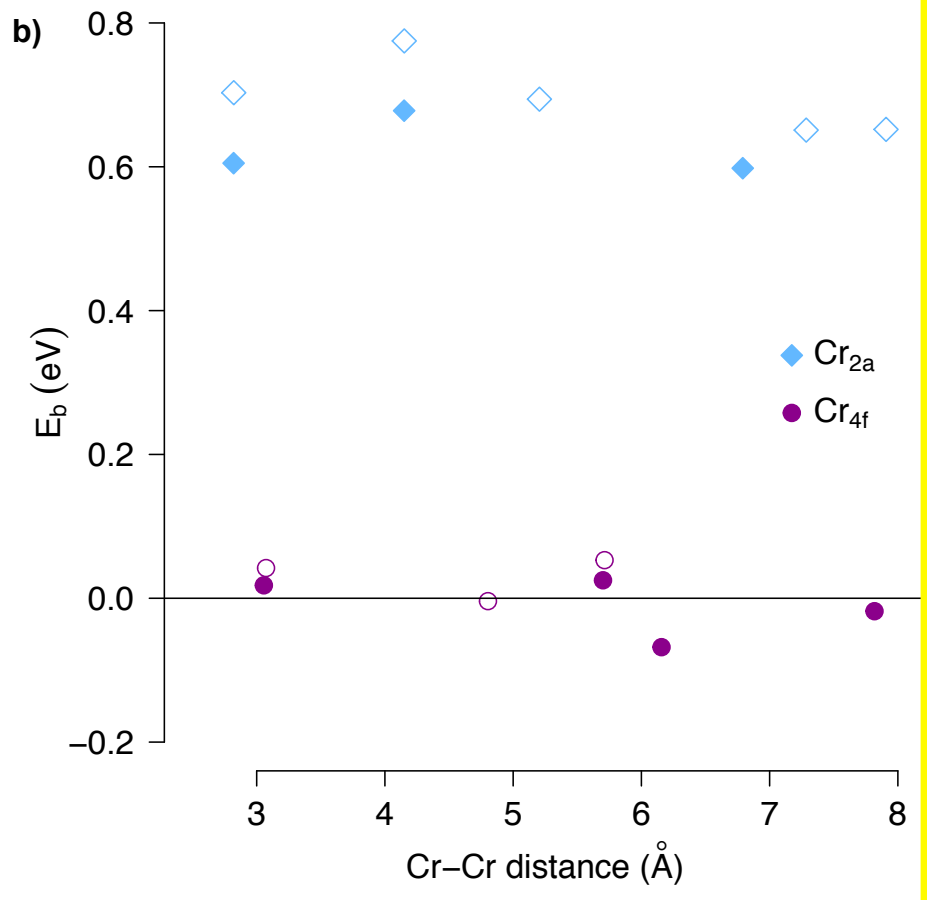

Figure 2 – **a)** SEM-EDX analyses of the *$Cr_5Ti_{95}$* sample. The image was taken in BSE mode. The area displayed, not representative of the whole sample, was selected as it contains more of the two minor phases relative to the rest of the sample. Determination of the phase compositions is described in the main text. **b)** Binding energy as a function of Cr-Cr distance of $2Cr_{Ti}$ clusters in $Ti_3AlC_2$. At least one of the two $Cr_{Ti}$ is on the less unfavorable *4f* site. Filled and hollow symbols were obtained from smaller supercells (108 atoms) and large supercells (192 atoms) respectively.

Accommodation of Cr in $Ti_3AlC_2$ was investigated at the atomic scale with DFT simulations, with the assumption that Cr would exclusively reside on the metallic M sites. This choice was made given prior DFT studies on this and similar systems [3,19] but also neutron diffraction studies such as that of Caspi et al. [39] which all agree the usual M, A and X elements are only present on their respective crystallographic sites in the absence of radiation damage. The solution energies associated with the substitution of Cr for Ti were calculated for the two distinct Wyckoff sites:

$$Cr_{(s)} + Ti_{Ti(4f)} \xrightarrow{1.28\ eV} Cr_{Ti(4f)} + Ti_{(s)} \quad (1)$$

$$Cr_{(s)} + Ti_{Ti(2a)} \xrightarrow{1.91\ eV} Cr_{Ti(2a)} + Ti_{(s)} \quad (2)$$

where Kröger-Vink notation is used the denote the substitution defect [40]. In both cases the solution energy is positive (unfavorable) suggesting that Cr should exhibit only limited solubility in $Ti_3AlC_2$, in agreement with the experimental findings presented above. Notably, substitution onto the *4f* site is 0.63 eV less unfavorable compared to substitution on the *2a* site. Consequently, if small (dilute) amounts of Cr were added to $Ti_3AlC_2$, according to the Boltzmann distribution, 98.5 % of the total Cr content should reside on the *4f* site at the synthesis temperature of 1450ºC, with the ratio increasing to 99.2 % at the sintering temperature of 1220ºC.

The relaxation volume associated with the incorporation of a single Cr atom, as defined in [41], was calculated as -6.508 Å$^3$, in a 3x3x1 supercell containing 108 atoms (total supercell volume = 1378.15 Å$^3$). This equates to a change in volume of −0.51% for every addition of 1 at% Cr. Equivalently, at the solubility limit of 2 at% Cr, as measured by EDX above, the predicted volume change from $Ti_3AlC_2$ to $(Cr_{0.02}Ti_{0.98})_3AlC_2$ is only 1.02 vol%. This relatively small change confirms that standard powder XRD is unable to prove or disprove Cr presence in $Ti_3AlC_2$ at such low substitution rate, given the wider range of existing unit cell parameter values for $Ti_3AlC_2$ [34-38].

To investigate whether the Cr solubility was dependent on Cr concentration, defect clusters containing two Cr atoms were considered, in which at least one Cr atom occupies the more favorable *4f* site. Figure 2 (b) shows the binding energy associated with the formation of a $\{Cr_{Ti}: Cr_{Ti}\}$ cluster, following equation 3

$$E_b = E^f{}_{2\{Cr_{Ti}\}} - 2\,E^f{}_{Cr_{4f}} \tag{3}$$

where $E^f$ is the formation energy of the defect. A negative bonding energy indicates increased stability against dilute defects. Clusters where both Cr atoms occupy *4f* sites show small but negative binding energies. On the other hand, the clusters where one Cr atom occupies the *2a* site are unstable by 0.6-0.7 eV, which is similar to the energy difference between the accommodation of Cr in *2a* vs *4f* sites in the dilute cases. This indicates that Cr exhibits limited solubility in $Ti_3AlC_2$ even when defect clusters are considered. It also indicate that Cr prefers to occupy the *4f* site at low and higher concentrations alike, and will therefore tend towards the formation of a layered structure where all the $Cr_{Ti}$ allowed to enter $Ti_3AlC_2$ resides on *4f* sites, in line with the experimental observation of ordered $(Cr_{2/3}Ti_{1/3})_3AlC_2$ [24].

## 4. Ordering and non-stoichiometry of $(Cr_{2/3}Ti_{1/3})_3AlC_2$.

$(Cr_{2/3}Ti_{1/3})_3AlC_2$ was first obtained and reported by Liu et al. in 2014 [22,24] and in parallel synthesized by our research group and reported a year later [23]. We agree that the stable polymorph at room temperature of this MAX phase exhibits the α structure. After a thorough investigation by Raman spectroscopy, TEM, XRD and especially neutron diffraction (ND), Liu et al. [24] further found experimentally that α-$(Cr_{2/3}Ti_{1/3})_3AlC_2$ has an ordered structure since the Cr atoms are, in the limit of sensitivity of ND, all located on the *4f* sites while the Ti atoms are only present on the *2a* sites. This was corroborated by Dahlqvist and Rosen [19] who calculated by DFT the theoretical temperature of transition from an ordered to a disordered state to be over 3300°C, thereby inaccessible with conventional powder processing techniques. An interesting point also found in previous studies [22,23] concerns synthesis attempts of compositions close to $(Cr_{2/3}Ti_{1/3})_3AlC_2$, which are $(Cr_{3/4}Ti_{1/4})_2AlC$, $(Cr_{1/2}Ti_{1/2})_3AlC_2$, $(Cr_{4/7}Ti_{3/7})_3AlC_2$ and $(Cr_{5/6}Ti_{1/6})_3AlC_2$: in all cases a *312* MAX phase with diffraction patterns and lattice parameters in line with that of "pure" $(Cr_{2/3}Ti_{1/3})_3AlC_2$ samples was formed along with either TiC (when $(Cr_{1/2}Ti_{1/2})_3AlC_2$ and $(Cr_{4/7}Ti_{3/7})_3AlC_2$ are originally targeted) or $CrC_x$ + $Cr_2AlC$ (for $(Cr_{5/6}Ti_{1/6})_3AlC_2$). This thus signifies that $(Cr_{2/3}Ti_{1/3})_3AlC_2$ prefers to segregate excess Cr or Ti in secondary phases rather than to deviate from its nominal 2/3 Cr + 1/3 Ti ratio. The unit cell parameters determined for the *312* MAX phases in "$(Cr_{3/4}Ti_{1/4})_2AlC$", "$(Cr_{1/2}Ti_{1/2})_3AlC_2$" and "$(Cr_{2/3}Ti_{1/3})_3AlC_2$" samples, given in Ref. [23], are very close one to another − typically within associated errors. This indicates that if a deviation of stoichiometry in $(Cr_{2/3}Ti_{1/3})_3AlC_2$ occurs when the reactants contain an inadequate ratio of Cr/Ti/Al/C, the deviation is low enough that the resulting variation in lattice parameters measured by XRD falls within the range of reported XRD values for the lattice parameters of $(Cr_{2/3}Ti_{1/3})_3AlC_2$ ($2.921 \leq a \leq 2.935$, $17.878 \leq c \leq 17.894$) [22,23]. Similarly, the neutron diffraction refinement carried out by Liu et al [24] did not show any sign of systematic non-stoichiometry in $(Cr_{2/3}Ti_{1/3})_3AlC_2$ (e.g. through the presence of vacancies). To try further investigating possible non-stoichiometry in $(Cr_{2/3}Ti_{1/3})_3AlC_2$, EDX was performed on the samples that we previously obtained when aiming at synthesizing $(Cr_{3/4}Ti_{1/4})_2AlC$, $(Cr_{1/2}Ti_{1/2})_3AlC_2$ and $(Cr_{2/3}Ti_{1/3})_3AlC_2$ thereafter labeled *211-Cr₃/₄Ti₁/₄*, *312-Cr₁/₂Ti₁/₂* and *312-Cr₂/₃Ti₁/₃*, respectively. Some characteristics of these samples syntheses are given in Table 1 for clarity.

Table 1 – $(Cr_{2/3}Ti_{1/3})_3AlC_2$-containing samples considered for EDX characterization.

| **Sample name** | **Targeted compound** | **Powder batch composition** | **Obtained phases** (and $Cr_{2/3}Ti_{1/3}AlC_2$ volume ratio) |
|---|---|---|---|
| *211-$Cr_{3/4}Ti_{1/4}$* | $(Cr_{0.75}Ti_{0.25})_2AlC$ | 1.5 Cr + 0.5 $TiH_2$ + 1.05 Al + 0.95 C | $Cr_{2/3}Ti_{1/3}AlC_2$ (40%) + TiC + $Al_8Cr_5$ + $AlCr_2$ |
| *312-$Cr_{1/2}Ti_{1/2}$* | $(Cr_{0.5}Ti_{0.5})_3AlC_2$ | 1.5 Cr + 1.5 $TiH_2$ + 1.05 Al + 1.9 C | $Cr_{2/3}Ti_{1/3}AlC_2$ (50%) + TiC + $Al_8Cr_5$ + $AlCr_2$ |
| *312-$Cr_{2/3}Ti_{1/3}$* | $(Cr_{2/3}Ti_{1/3})_3AlC_2$ | 2 Cr + 1 $TiH_2$ + 1.05 Al + 1.9 C | $Cr_{2/3}Ti_{1/3}AlC_2$ (95%) + TiC + $Al_8Cr_5$ |

An example of EDX results is given in Figure 3 for *211-$Cr_{3/4}Ti_{1/4}$*. This SEM sample was analyzed in three separate occasions (8 months span time) and as detailed hereafter, the average quantifications from one session to another were found to be less than 1% different, suggesting the absence of a drifting of the EDX apparatus with time. In the example presented in Figure 3, we first note that the average Al/(Al+Ti+Cr) ratio for 6 measurements in the $(Cr_{2/3}Ti_{1/3})_3AlC_2$ grains is 24.7 ± 0.7 at%. The value increases to 25.0 ± 2.1 at% (standard deviation σ = 1.0%) when averaged over 24 measurements taken during the three sessions and over a number of different locations. These values compare very closely with the EDX measurements performed on the *312-$Cr_{1/2}Ti_{1/2}$* (25.4 ± 1.1 at%) and the *312-$Cr_{2/3}Ti_{1/3}$* (24.6 ± 1.5 at%) samples. The total average for the $(Cr_{2/3}Ti_{1/3})_3AlC_2$ phase across all samples remains 25.0 ± 2.1 at% (σ = 0.9%). Amongst MAX phases, one can easily find works where non-stoichiometries were measured [1,42-47] or suggested according to DFT calculations [20,47-51]. As in most if not all cases an X or A elemental deficiency was reported (e.g. $Ti_4AlN_{2.9}$ [44,45], $Ti_3Ge_{0.8}C_2$ [42]) it appears reasonable to make the assumption that M sites in $(Cr_{2/3}Ti_{1/3})_3AlC_2$ are fully occupied and that this can serve as the necessary reference for stoichiometry determination. Thereby, as the Al/(Al+Ti+Cr) ratio stays very close to 25 at% throughout the measurement sessions and the samples, it can be concluded that EDX, in the limit of its sensitivity, was not able to evidence any deficiency (or excess) in Al and the stoichiometry of Al in $(Cr_{2/3}Ti_{1/3})_3AlC_2$ is therefore expected to be exactly 1, regardless of the initial reactants stoichiometries.

Regarding the Cr/(Cr+Ti) ratio in *211-$Cr_{3/4}Ti_{1/4}$*, the results are presented following the same data treatment and considering the same EDX points selected for the previous analysis on Al content. The Cr/(Cr+Ti) ratio for the 6 data points given in Figure 3 is 65.4 ± 2.4%. The average over 24 points is 65.7 ± 2.7% (σ = 1.6%). For the *312-$Cr_{1/2}Ti_{1/2}$* and *312-$Cr_{2/3}Ti_{1/3}$* samples, the average (Cr/(Cr+Ti) ratio are of 65.3 ± 3.4% (σ = 2.0%) and 65.8% ± 1.4% (σ = 0.6%), respectively; this leads to an average value of 65.7 ± 3.3% (σ = 1.5%) over all three samples. Once again the measured ratio does not vary much from a measurement session to another, proving the stability over time of the SEM/EDX apparatus, but more importantly Cr/(Cr+Ti) does not vary from one sample to another, despite differences in initial powders stoichiometries (excess Ti in the *312-*$Cr_{1/2}Ti_{1/2}$ sample and excess Cr in the *211-*$Cr_{3/4}Ti_{1/4}$ sample). Therefore it can be inferred that, within the limits of precision of our EDX characterization, the Cr/(Cr+Ti) ratio in $(Cr_{2/3}Ti_{1/3})_3AlC_2$ appears independent of the reactants stoichiometries. Besides one can also note that all samples show a Cr/(Cr+Ti) about 1% lesser than the theoretical and expected 2/3 (66.7%) value. Of course the scattering of data and the associated standard deviation are generally greater than 1%, but the fact that the deviation from the 2/3 theoretical value is for the 3 samples always falling on the same side indicates that if a real deviation exists, this EDX characterization suggests it to be in favor of a Ti excess over Cr, i.e. through the substitution of Ti for Cr onto *4f* Wyckoff sites, as it is calculated hereafter by DFT.

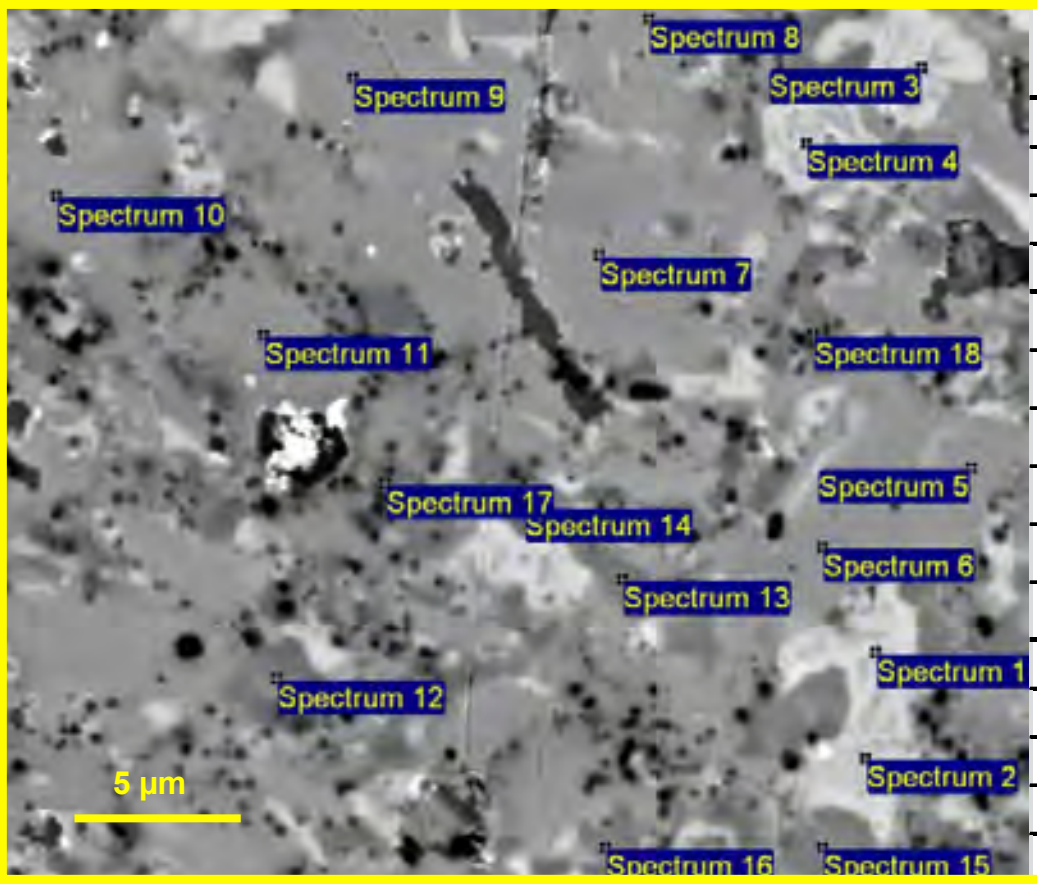

| Spectrum # | Al/(Al+Ti+Cr) (%) | Ti/(Al+Ti+Cr) (%) | Cr/(Al+Ti+Cr) (%) | Determined Composition |
|---|---|---|---|---|
| 1 | 35.8 | 3.0 | 61.2 | $AlCr_2$ |
| 2 | 35.7 | 3.0 | 61.3 | |
| 3 | 36.5 | 2.7 | 60.9 | |
| 4 | 34.4 | 2.0 | 63.7 | |
| 5 | 25.2 | 24.7 | 50.2 | $(Cr_{2/3}Ti_{1/3})_3AlC_2$<br>Al = 24.7%<br>Ti = 26.1%<br>Cr = 49.2% |
| 6 | 24.1 | 27.3 | 48.6 | |
| 7 | 24.4 | 26.4 | 49.3 | |
| 8 | 24.9 | 24.2 | 50.9 | |
| 9 | 24.6 | 27.3 | 48.1 | |
| 10 | 25.3 | 26.4 | 48.3 | |
| 13 | 50.4 | 5.9 | 43.6 | $Al_8Cr_5$ |
| 14 | 47.9 | 8.5 | 43.6 | |
| 15 | 49.2 | 9.0 | 41.8 | |
| 16 | 47.2 | 10.2 | 42.6 | |
| 18 | 45.3 | 17.4 | 37.3 | |

Figure 3 – SEM-EDX analyses of the *211-*$Cr_{3/4}Ti_{1/4}$ sample. The image was taken in BSE mode.

For DFT investigations, the α-$(Cr_{2/3}Ti_{1/3})_3AlC_2$ structure was first simulated by ordering the Cr atoms on the *4f* sites and the Ti atoms on the *2a* sites (Figure 1), in line with DFT results from Dahlqvist and Rosen [7], experimental findings of Liu et al. [24] and observations of such ordering in similar MAX phases [39,52]. The formation of energy of $(Cr_{2/3}Ti_{1/3})_3AlC_2$ and its ternary end members are presented, in Eqs. 3-5:

$$3\mathrm{Ti}_{(hcp)} + \mathrm{Al}_{(fcc)} + \mathrm{C}_{(2h)} \xrightarrow{-4.68\ \mathrm{eV}} \mathrm{Ti_3AlC_2} \qquad \tfrac{1}{2}\mathrm{Ti_2AlC} + \tfrac{1}{2}\mathrm{Ti_4AlC_3} \xrightarrow{-0.078\ \mathrm{eV}} \mathrm{Ti_3AlC_2} \tag{3a,b}$$

$$2\mathrm{Cr}_{(bcc)} + \mathrm{Ti}_{(hcp)} + \mathrm{Al}_{(fcc)} + \mathrm{C}_{(2h)} \xrightarrow{-2.31\ \mathrm{eV}} \mathrm{Cr_2TiAlC_2} \qquad \mathrm{Cr_2AlC} + \mathrm{TiC} \xrightarrow{-0.081\ \mathrm{eV}} \mathrm{Cr_2TiAlC_2} \tag{4a,b}$$

$$3\mathrm{Cr}_{(bcc)} + \mathrm{Al}_{(fcc)} + \mathrm{C}_{(2h)} \xrightarrow{-0.26\ \mathrm{eV}} \mathrm{Cr_3AlC_2} \qquad \mathrm{Cr_2AlC} + \tfrac{1}{3}\mathrm{Cr_3C_2} + \tfrac{1}{3}\mathrm{C}_{(2h)} \xrightarrow{+0.41\ \mathrm{eV}} \mathrm{Cr_3AlC_2} \tag{5a,b}$$

Here we consider both the formation energy from standard state (Eqs. 3a, 4a, 5a), and the reaction energy from the most competing phases, as identified by Dahlqvist and Rosen [19,20] (Eqs. 3b, 4b and 5b). $(Cr_{2/3}Ti_{1/3})_3AlC_2$ exhibits the lowest reaction energy when considering the most competing phases.

It has been shown that $(Cr_{2/3}Ti_{1/3})_3AlC_2$ exhibits multiple metastable spin states [19,53], therefore all previously reported spin states were considered and the reaction energies of equation 4b are presented in Table 2. In agreement with Dahlqvist and Rosen [19], we find that the in-plane antiferromagnetic ordering type XX (following the nomenclature of Wang *et al.* [53]) is the most favorable spin state, closely followed by simple ferromagnetic ordering with a difference of only 1meV. Note that this energy difference is much below the expected accuracy of point defect calculations within DFT (see methodology section, and the effect of on-site Coulomb correction discussed by Wang *et al.* [53]), therefore all defect calculations were performed with the ferromagnetic ordering only, thus allowing greater flexibility on the size of supercell adopted.

Table 2 – Relative stability of $(Cr_{2/3}Ti_{1/3})_3AlC_2$ under different spin configurations.

| Spin order | Supercell size | Eq. 4b (meV) |
|---|---|---|
| NM | unit cell | 40 |
| FM | unit cell | -80 |
| AFM-X | unit cell | -62 |
| AFM-1 | unit cell | 21 |
| AFM-A | unit cell | 0 |
| AFM-FF | 2x1x1 | -62 |
| AFM-XX | 2x1x1 | -81 |
| AFM-11 | 2x1x1 | 33 |
| AFM-AA | 2x1x1 | 36 |

As discussed in the previous section, if the phase forms by progressive additions of Cr into the $Ti_3AlC_2$ structure, then it is expected to retain the ordered, layered structure of $(Cr_{2/3}Ti_{1/3})_3AlC_2$. However, to investigate the driving force for ordering of $(Cr_{2/3}Ti_{1/3})_3AlC_2$, the formation energy of anti-site defects (i.e. swapping the positions of a Cr and a Ti atom to form a $\{Ti_{Cr}:Cr_{Ti}\}$ defect pair) is considered, following the method outlined in [54]. Six anti-site defect configurations were considered: three configurations where the swapping atoms reside in neighbouring planes (part of the same carbide layer) and three configurations where the defects are separated by the Al layer. In all cases, the formation energy was unfavourable at 0.82 ± 0.06 eV, confirming that there is a significant preference for ordered layering of Cr and Ti atoms at this composition.

Interestingly, similarly to the $\{Cr_{Ti}:Cr_{Ti}\}$ defects clusters in $Ti_3AlC_2$ discussed earlier, the formation energies of the anti-site clusters were found to be largely insensitive to the distance or location of the defect pair (standard deviation of only 0.06 eV). Furthermore, the formation energy for dilute non-interacting defects (i.e. $Ti_{Cr}$ + $Cr_{Ti}$ in separate simulations) was calculated at 0.80 eV, which is within the uncertainty of the formation energy of the bound defects. This suggests that there is no driving force for either clustering or separation of anti-site defects if these were forcefully introduced in the system (e.g. through radiation damage).

The ability of the structure to accommodate more (or less) Cr than exactly 2/3 of the total M sites can be calculated from the formation energy of individual $Cr_{Ti}$ (or $Ti_{Cr}$) defects in the ordered $(Cr_{2/3}Ti_{1/3})_3AlC_2$ structure. The formation energy of the defect is sensitive to the choice of reference phase, i.e. what is the chemical form of the local reservoirs of Ti and Cr atoms. Here we consider six possible scenarios where the reference phases were taken (eq. 6) from standard state (i.e. metallic Ti and Cr), (eq. 7) from the parent 312-MAX phases, and (eq. 8) from the lower order 211-MAX phases.

$$\mathrm{Ti_{(s)}} + \mathrm{Cr_{Cr}} \xrightarrow{-1.10} \mathrm{Ti_{Cr}} + \mathrm{Cr_{(s)}} \qquad \mathrm{Cr_{(s)}} + \mathrm{Ti_{Ti}} \xrightarrow{1.90} \mathrm{Cr_{Ti}} + \mathrm{Ti_{(s)}} \tag{6a,b}$$

$$\tfrac{1}{3}\,\mathrm{Ti_3AlC_{2(s)}} + \mathrm{Cr_{Cr}} \xrightarrow{0.37} \mathrm{Ti_{Cr}} + \tfrac{1}{3}\,\mathrm{Cr_3AlC_{2(s)}} \qquad \tfrac{1}{3}\,\mathrm{Cr_3AlC_{2(s)}} + \mathrm{Ti_{Ti}} \xrightarrow{0.43} \mathrm{Cr_{Ti}} + \tfrac{1}{3}\,\mathrm{Ti_3AlC_{2(s)}} \tag{7a,b}$$

$$\tfrac{1}{2}\,\mathrm{Ti_2AlC_{(s)}} + \mathrm{Cr_{Cr}} \xrightarrow{0.04} \mathrm{Ti_{Cr}} + \tfrac{1}{2}\,\mathrm{Cr_2AlC_{(s)}} \qquad \tfrac{1}{2}\,\mathrm{Cr_2AlC_{(s)}} + \mathrm{Ti_{Ti}} \xrightarrow{0.76} \mathrm{Cr_{Ti}} + \tfrac{1}{2}\,\mathrm{Ti_2AlC_{(s)}} \tag{8a,b}$$

The values reported do not consider all possible phases that may form, only the simplest, most intuitive, cases where reactants and products are the constituent metals or ternary MAX phases or carbides. A comprehensive analysis of all the possible binary, ternary and quaternary phases that may form, is clearly beyond the scope of this investigation, and conclusions may be inferred without such considerations. For instance, since the formation energy for accommodating excess Cr (Eqs. 6b, 7b and 8b) is unfavourable in all scenarios, then it is clear that considering additional (more stable) reactants may only render $Cr_{Ti}$ accommodation even more unfavourable, thus $(Cr_{2/3}Ti_{1/3})_3AlC_2$ is expected to show no meaningful deviation of stoichiometry in the case of Cr excess. These findings are in agreement with the results of Dahlqvist and Rosen [19], where they simulated the compositional range $Ti_{3-x}Cr_xAlC_2$ where $x$={0,0.5,1,1.5,2,2.5,3} (equivalent to concentration steps of 8.33 at%), whereas the current work considered dilute deviations from stoichiometry of ~0.925 at% (1 atom replaced in 108-atom supcercells). Regarding Ti excess (Eqs. 6a, 7a and 8a), on the other hand, the reaction energy varies from highly favorable (−1.10 eV) in the case of metal reactants (unlikely in the presence of C) to highly unfavorable (0.37 eV) in the case of a hypothetical MAX phase $Cr_3AlC_2$ product. Since $Cr_3AlC_2$ is known not to form, any binary or ternary compound that forms in its place *must* have lower formation energy, thus reducing the reaction energy for accommodation of $Ti_{Cr}$. Nevertheless, irrespective of reactants and products, the accommodation of excess Ti is consistently more favourable than the accommodation of excess Cr, thus suggesting that if any non-stoichiometry may be accommodated in $(Cr_{2/3}Ti_{1/3})_3AlC_2$, it will necessarily be towards Ti excess.

## 5. Factors contributing to the stability of $(Cr_{2/3}Ti_{1/3})_3AlC_2$.

Calculated lattice parameters of the $(Cr_xTi_{1-x})_3AlC_2$ series are presented in Figure 4, including, for the purpose of comparison, hypothetical ordered phases $(Cr_{1/3}Ti_{2/3})_3AlC_2$ and $Cr_3AlC_2$ that are known not to form experimentally. Incidentally those phases that are thought to be unstable also proved difficult to converge at the computational level, suggesting that perhaps at the root of their instability is the electron distribution. Only the lowest energy magnetic configuration is reported for each composition in Figure 4 as these best represent real-life materials. However, changes in magnetic ordering are known to affect the lattice parameters of these materials significantly, as thoroughly investigated by Wang et al. [53]. Notably, the change in lattice parameters between ferromagnetic and non-magnetic $Cr_3AlC_2$ was of the same order as the change in lattice parameter between $Ti_3AlC_2$ and $Cr_3AlC_2$. The calculated lattice parameter of $Ti_3AlC_2$ ($a$ = 3.078 Å, $c$ = 18.654 Å) fall within the range of experimental values from literature (3.065 Å ≤ $a$ ≤ 3.0786 Å and 18.487 Å ≤ $c$ ≤18.73 Å [34-38]). For $(Cr_{2/3}Ti_{1/3})_3AlC_2$ the calculated lattice parameters vary significantly depending on the spin ordering, however for the two most stable spin states the lattice parameters were $a^{FM}$ = 2.930 Å, $c^{FM}$ = 17.830 Å and $a^{AFM\text{-}XX}$ = 2.921 Å, $c^{AFM\text{-}XX}$ = 17.874 Å. These are also within or very close to the range of experimental values found in the literature: 2.921 Å ≤ $a$ ≤ 2.935 Å, 17.878 Å ≤ $c$ ≤ 17.894 Å [23,24]. It is well established that the choice of GGA exchange-correlation functional may lead to an overestimation or underestimation of the lattice parameter for most material systems. However, it is also acknowledged that the error is generally consistent within a system series, therefore the prediction of lattice parameter change due to a change in composition is typically highly accurate with DFT-GGA.

$(Cr_{2/3}Ti_{1/3})_3AlC_2$ exhibits a significant deviation from Vegard's law, resulting in an overall reduction in volume compared to the linear trend and, surprisingly, an exceptional increase in c/a ratio above that of any other ternary or quaternary *312*-MAX phase in the series. This is a strong indication that the nature of the in-plane and out-of-plane bonds is different at this specific composition. To investigate this further, the electronic densities of states (DOS) were calculated for the four phases and are presented in Figure 5.

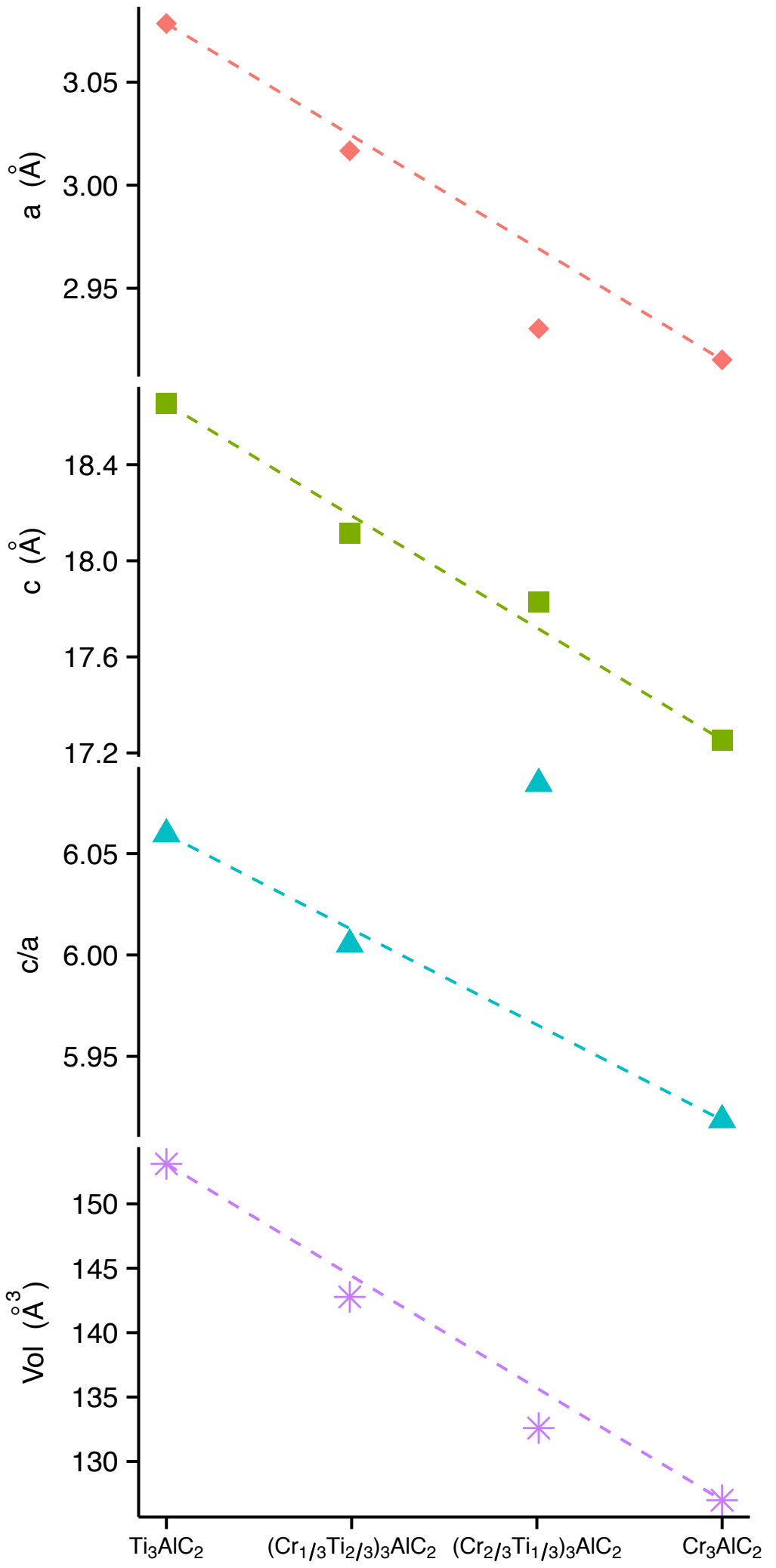

Figure 4 – Variation of lattice parameters *a*, *c*, *c/a* ratio and unit cell volume as a function of Cr content in $(Cr_xTi_{1-x})_3AlC_2$.

Addition of Cr creates sharp peaks in the energy region near the Fermi level (Figure 5), which are typically indicative of localized states. Furthermore, addition of Ti or Cr beyond the 2/3:1/3 ratio causes spin polarization of the structure as indicated by the asymmetric shape of the DOS of $(Cr_{1/3}Ti_{2/3})_3AlC_2$ and $Cr_3AlC_2$. The DOS of the stable phases, $(Cr_{2/3}Ti_{1/3})_3AlC_2$ and $Ti_3AlC_2$, is further expanded into the individual orbital contributions in Figure 6. It is evident that spin polarization of the quaternary MAX phase is not limited to the Cr atoms (where it is most pronounced), but it is also observed in the projected DOS of all other atomic species. Furthermore Al and C exhibit changes in densities around the Fermi level, with new narrow bands forming in the spin-up component of the DOS at the same energies as the most pronounced Cr spin-up bands, suggesting that there are (partly) covalent Al-Cr and C-Cr bonds in $(Cr_{2/3}Ti_{1/3})_3AlC_2$. To further investigate this, Mulliken population analysis was performed (Table 3) and the charge density difference (with respect to the sum of individual atomic charge densities) was extracted from the relaxed lattices (Figure 7). The results are shown for the ferromagnetic configuration only, but the AFM-XX structure yielded population analysis results within 2% of the FM values.

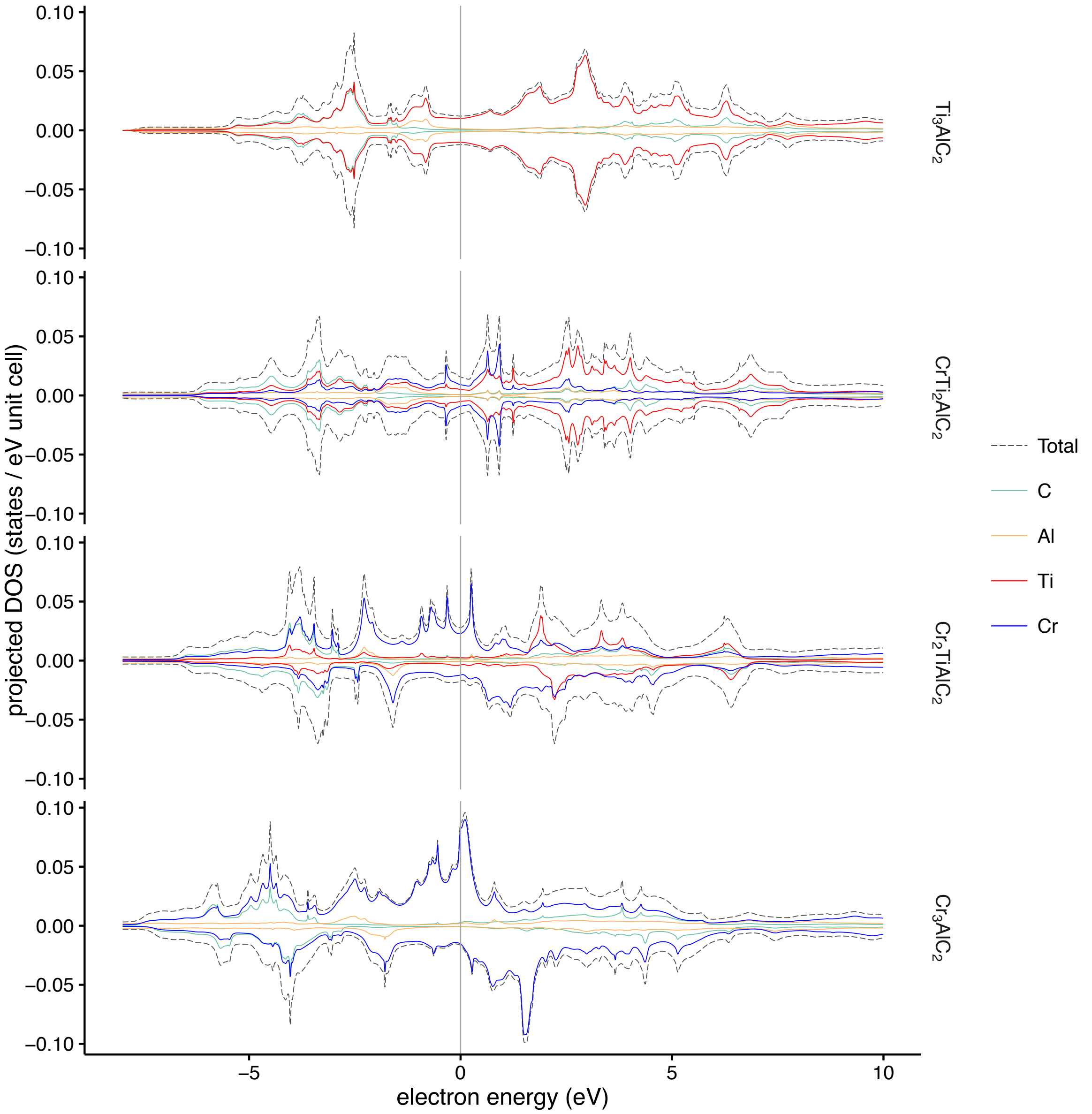

Figure 5 – Electronic density of states, aligned with respect to the Fermi level, of the $(Cr,Ti)_3AlC_2$ series.

The *4f* site – occupied by Cr in $(Cr_{2/3}Ti_{1/3})_3AlC_2$ – links the carbide layer with the aluminum layer, whilst the *2c* site – occupied only by Ti – is surrounded exclusively by C atoms (see Figure 1). Consequently, it seems sensible to attribute the increase in stability of the $(Cr_{2/3}Ti_{1/3})_3AlC_2$ phase to Al-Cr interactions. However, the population analysis shows that the electron density of the Al-Cr bond in $(Cr_{2/3}Ti_{1/3})_3AlC_2$ is reduced by ~20 vol% compared to the Al-Ti bond in $Ti_3AlC_2$. And whilst the Al-Cr bond is 19.3 pm shorter than the Al-Ti bond, which typically is an indication of a stronger bond, the difference in bond length is entirely accounted for by the smaller atomic radius of Cr compared to Ti (124.9 pm vs 144.8 pm) [55].

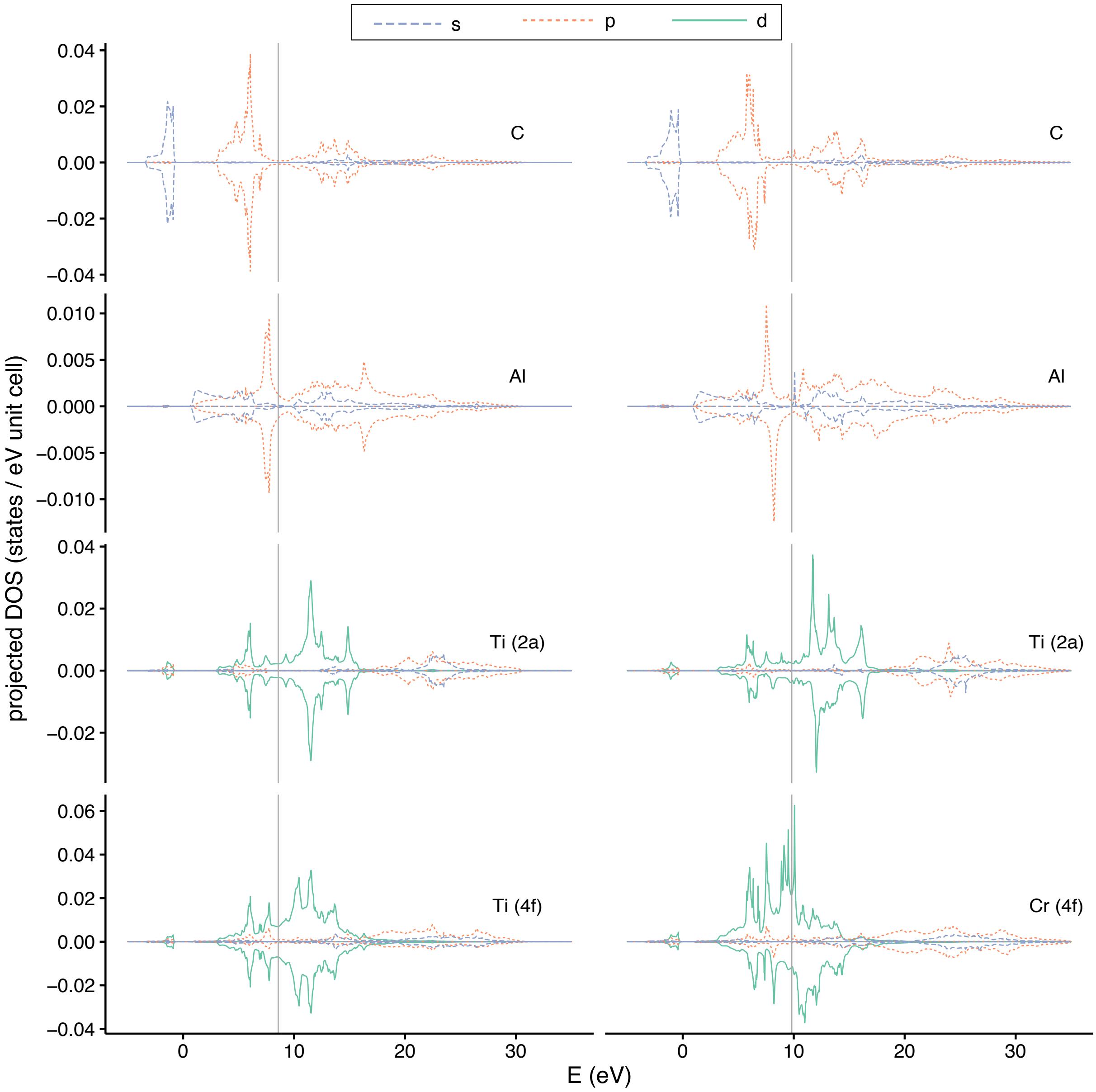

Figure 6 – Electronic DOS of $Ti_3AlC_2$ (left) and $(Cr_{2/3}Ti_{1/3})_3AlC_2$ (right). Vertical lines represent the Fermi level.

Table 3 – Mulliken bond and population analysis of $(Cr_{2/3}Ti_{1/3})_3AlC_2$ (top) and $Ti_3AlC_2$ (bottom). The last three columns refers to in-plane interactions.

| $\mathbf{(Cr_{2/3}Ti_{1/3})_3AlC_2}$ | **C-Ti(*2a*)** | **C-Cr(*4f*)** | **Al-Cr(*4f*)** | **Ti(*2a*)-Cr(*4f*)** | **Cr(*4f*)-Cr(*4f*)** | **Ti(*2a*)-Ti(*2a*)** | **Al-Al** |
|---|---|---|---|---|---|---|---|
| Bond length (pm) | 215 | 197 | 271 | 289 | 293 | 293 | 293 |
| Population | 0.65 | 1.35 | 0.39 | −0.84 | −0.22 | 0.25 | 0.57 |
| $\mathbf{Ti_3AlC_2}$ | **C-Ti(*2a*)** | **C-Ti(*4f*)** | **Al-Ti(*4f*)** | **Ti(*2a*)-Ti(*4f*)** | **Ti(*4f*)-Ti(*4f*)** | **Ti(*2a*)-Ti(*2a*)** | **Al-Al** |
| Bond length (pm) | 220 | 208 | 290 | 296 | 308 | 308 | 308 |
| Population | 0.88 | 1.17 | 0.48 | −0.55 | 0.08 | −0.05 | 0.42 |

On the other hand, a higher localization of electron is observed around the *4f* Cr atom compared to the *4f* Ti atoms in $Ti_3AlC_2$ (Figure 7). A corresponding increase in electron density of ~15 vol% is found around the C-Cr bond compared to the same sites in $Ti_3AlC_2$. At the same time, the C-Ti and Al-Cr bonds are markedly depopulated. This suggests that the increase in stability of the $(Cr_{2/3}Ti_{1/3})_3AlC_2$ is largely due to the C-Cr interaction, despite the preferential location of Cr atoms (bridging Al layers and carbide layers). As a consequence of the redistribution of electrons, the local electron density in and around the Al layer is increased (~36 vol%), which may also play a role in the stability of layered $(Cr_{2/3}Ti_{1/3})_3AlC_2$. Dahlqvist and Rosen [19] suggested that the addition of M elements with higher electronegativity than Ti (such as Mo in their paper and Cr in this paper) near the Al layer causes the Al atoms to become more positively charged, thus reducing the number of electrons available for the Al-Al anti-bond. This is in contrast to our finding that the local (relative) density around Al atoms increases. Nevertheless, Figure 6 shows that the DOS of the Al *p* orbital in $(Cr_{2/3}Ti_{1/3})_3AlC_2$ is shifted towards lower energies compared to $Ti_3AlC_2$, suggesting an energetically more favourable bond, in line with the findings of Dahlqvist and Rosen [19]. This is corroborated by a subtle change in the electron density surrounding the Al atom: the in-plane density maximum between Al atoms in $Ti_3AlC_2$ splits into two (stronger) maxima in $(Cr_{2/3}Ti_{1/3})_3AlC_2$, one above and one below the Al layer. Finally, it is observe that the in-plane M-M interactions also exhibit a significant charge transfer when Cr is included. In $Ti_3AlC_2$, the basal plane (*2a-2a*) Ti-Ti bonds are nearly identically populated to the in-plane (*4f-4f*) Ti-Ti bonds. However, in $(Cr_{2/3}Ti_{1/3})_3AlC_2$ the in-plane Cr-Cr bond (*4f-4f*) is markedly depopulated in favour of the Ti-Ti basal bond (*2a-2a*). This may be a further source of increased stability for the Ti carbide layers in the quaternary Cr-Ti-Al-C MAX phase.

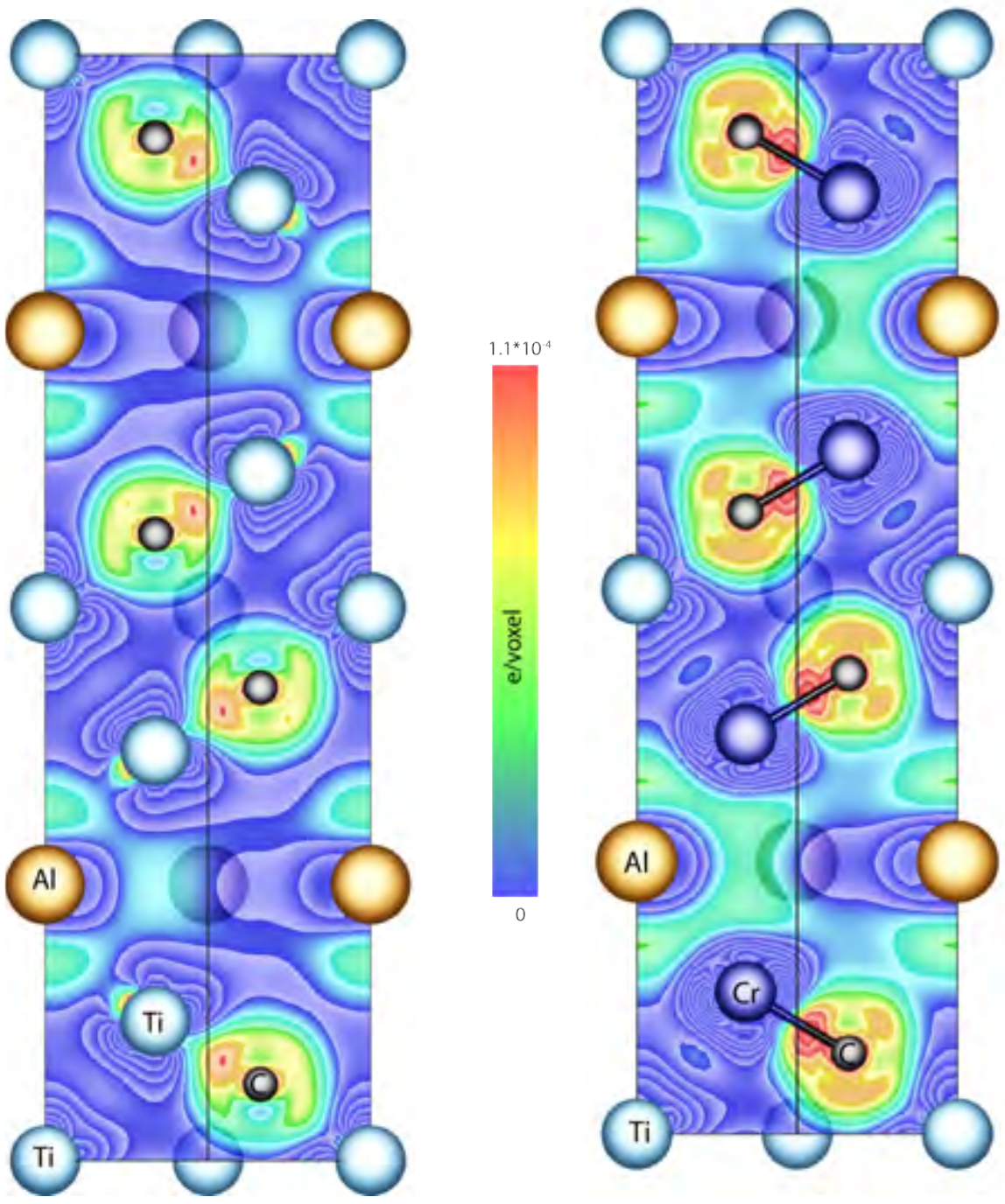

Figure 7 – (110) planar section of the charge density difference of $Ti_3AlC_2$ (left) and $(Cr_{2/3}Ti_{1/3})_3AlC_2$ (right). Units are electrons/voxel where the sum over all unit-cell space is equal to the total number of electrons in the unit cell.

It is believed that similar redistribution of bond charges might be at the root of stability of other $(M,M')_{n+1}AX_n$ MAX phases recently reported. Anasori et al. [52] reported a quaternary $(Mo,Ti)_3AlC_2$ with the same layered ordering and M/(M+Ti) ratio as $(Cr_{2/3}Ti_{1/3})_3AlC_2$. The analogy with the Cr-Ti system continues since, like $Cr_3AlC_2$, $Mo_3AlC_2$ is known not to form. Interestingly Zhou *et al.* [56] and Caspi *et al.* [39] reported stable $(Cr_{0.5},V_{0.5})_{n+1}AlC_n$ MAX phases with n=1,2,3. This behavior is distinctly different from that of the Cr-Ti-Al-C and Mo-Ti-Al-C systems, where stable phases are observed only at the specific M/(M+Ti) ratios that allow layered ordering of M and Ti atoms (2/3 in the case of *312*-MAX phases). In fact, the lack of a stable solid solution $(Cr,Ti)_2AlC$ [23] may be explained in terms of the inability of *211*-MAX phases (i.e. n=1) to achieve such layer ordering at any Cr/(Cr+Ti) ratio. The fact that $(Cr_{0.5},V_{0.5})_{n+1}AlC_n$ may be achieved even for n=1, suggests that the layered ordering of M atoms is less important to the phase stability in the V-Cr-Al-C system. This is further corroborated by the recent results of Hamlin *et al.* [57], where $(Cr_{0.5},V_{0.5})_{n+1}AlC_n$ MAX phases (n=1,2,3) were synthesized with Cr/(Cr+V) ratios of 1/4, 1/2 and 3/4 (the latter being unstable only for n=3). Nevertheless, a strong tendency towards ordering is observed for n=2 and n=3, as reported by the neutron diffraction data [39]. The results of the current work, together with those of Caspi *et al.* [39], suggest that if the stoichiometry of the compounds was adjusted to reflect the multiplicity of the different Wyckoff sites (i.e. Cr/(Cr+V) = 2/3 for n=2), then a V-Cr-Al-C *312*-MAX phase with layered ordering of Cr and V atoms may be obtained, and this may have improved stability. Nevertheless, the ability of $(Cr,V)_{n+1}AlC_n$ to accommodate disorder on the M site suggests that perhaps it may be more radiation tolerant than MAX phases such as $(Cr_{2/3}Ti_{1/3})_3AlC_2$ where ordering is a pre-requisite for stability.

## 6. Conclusions

Through a combination of DFT simulations and experimental synthesis, it was found that:

- Cr exhibits limited solubility in $Ti_3AlC_2$ (~2 at%) and that all the Cr is accommodated on the *4f* Wyckoff site, i.e. bridging the Al layer to the carbide layer.
- Larger amounts of Cr (Cr/(Cr+Ti) ratio ≥ 1/2) enable formation of $(Cr_{2/3}Ti_{1/3})_3AlC_2$, which is confirmed to be a fully ordered structure with Cr atoms on the *4f* sites and Ti atoms on the *2a* site.
- Deviation from stoichiometry of $(Cr_{2/3}Ti_{1/3})_3AlC_2$ is predicted and measured to be negligible in the case of Cr-excess but the results are inconclusive regarding Ti-excess, where if deviation occurs it is expected to be mediated by $Ti_{Cr}$ substitutions. EDX also found that the ratio Al/(Al+Cr+Ti) in $(Cr_{2/3}Ti_{1/3})_3AlC_2$ does not deviate significantly from the expected 1/4 value suggesting that the A site is fully occupied by Al.
- $(Cr_{2/3}Ti_{1/3})_3AlC_2$ has a smaller volume (132.6 Å$^3$ simulated, 133.2±0.2 Å$^3$ measured) and surprisingly larger *c/a* ratio (6.084 simulated, 6.101±0.003 measured) than the other *312*-MAX phases in the series.
- Increase in stability is primarily attributed to the formation of favourable Cr-C bond despite the preferential location of Cr (bridging Al and carbide layers), though magnetization may also play an important role, as may the redistribution of charge in the Al layer.
- Other $(M,M')_{n+1}AX_n$ MAX phases with layered ordering of M on *2a* sites and M' on *4f* sites are expected to exhibit increased stability due to the same factors (point above).

## 7. Acknowledgments

P.A.B. would like to thank J. N. Hart for the fruitful discussions and ANSTO and the Tyree foundation for financial support. Computational resources were provided by the Imperial College HPC service, the Australian National Computational Infrastructure supported by Intersect Australia Ltd (Raijin and Orange), and the Multi-modal Australian ScienceS Imaging and Visualisation Environment (MASSIVE). D.H. is grateful to S. Grasso and B. Milsom of Nanoforce lab, Queen Mary University, London for their help with powder preparation and SPS. This work was partially funded by XMat and CAFFE research projects (EPSRC grants EP/K008749/1 and EP/M018563/1, respectively).

# Supplementary materials – Experimental and DFT investigation of $(Cr,Ti)_3AlC_2$ MAX phase stability.

Patrick A. Burr*[,a], Denis Horlait[b,c], William E. Lee[c]

[a] School of EE&T, University of New South Wales, Kensington, NSW, 2052, Australia.
[b] CNRS/IN2P3 and University of Bordeaux, Centre d'Etudes Nucléaires de Bordeaux-Gradignan, UMR 5797, Chemin du Solarium, 33175 Gradignan, France.
[c] Centre for Nuclear Engineering (CNE) & Department of Materials, Imperial College London, SW7 2AZ, UK
* Corresponding author: p.burr@unsw.edu.au.

**Table S1 – Total energy of DFT simulations – unit cells**

| Compound | Polymorph | Atoms | Magnetic order | Energy (eV) |
|---|---|---|---|---|
| $Cr_3AlC_2$ | α | 12 | NM | -15694.178 |
| $Cr_3AlC_2$ | α | 12 | FM | -15694.407 |
| $Cr_3AlC_2$ | β | 12 | NM | -15693.993 |
| $Cr_3AlC_2$ | β | 12 | FM | -15694.222 |
| $Cr_2TiAlC_2$ | α | 12 | NM | -13963.279 |
| $Cr_2TiAlC_2$ | α | 12 | FM | -13963.518 |
| $Cr_2TiAlC_2$ | α | 12 | AFM-X* | -13963.483 |
| $Cr_2TiAlC_2$ | α | 12 | AFM-1* | -13963.317 |
| $Cr_2TiAlC_2$ | α | 12 | AFM-A* | -13963.359 |
| $Cr_2TiAlC_2$ | α | 24 | AFM-FF* | -27926.966 |
| $Cr_2TiAlC_2$ | α | 24 | AFM-XX* | -27927.043 |
| $Cr_2TiAlC_2$ | α | 24 | AFM-11* | -27926.584 |
| $Cr_2TiAlC_2$ | α | 24 | AFM-AA* | -27926.575 |
| $Cr_2TiAlC_2$ | β | 12 | NM | -13963.090 |
| $Cr_2TiAlC_2$ | β | 12 | FM | -13963.294 |
| $CrTi_2AlC_2$ | α | 12 | NM | -12229.518 |
| $CrTi_2AlC_2$ | α | 12 | FM | *unstable* |
| $Ti_3AlC_2$ | α | 12 | NM | -10498.302 |
| $Ti_3AlC_2$ | β | 12 | NM | -10497.832 |
| $Cr_2AlC$ | α | 8 | NM | -10443.924 |
| $Cr_2AlC$ | α | 8 | FM | -10443.934 |
| $Cr_2AlC$ | β | 8 | NM | -10443.697 |
| $Cr_2AlC$ | β | 8 | FM | -10443.714 |
| $Ti_2AlC$ | α | 8 | NM | -6978.555 |
| $Ti_2AlC$ | β | 8 | NM | -6978.160 |
| Al | FCC | 1 | NM | -128.023 |
| C | *2h* | 4 | NM | -630.798 |
| Cr | BCC | 2 | AFM | -4935.678 |
| Ti | HCP | 2 | NM | -3200.703 |
| $Ti_4AlC_3$ | α | 16 | NM | -14017.7378 |
| $Cr_3C_2$ | – | 20 | NM | -30877.016 |
| TiC | – | 8 | NM | -7038.850 |

*anti-ferromagnetic configurations follow the nomenclature of Wang *et al*. [53].

**Table S2 – Total energy of DFT simulations – supercells**

| Compound | Defect | Composition | Atoms | Energy (eV) |
|---|---|---|---|---|
| $Cr_2TiAlC_2$ | – | $Cr_{36}Ti_{18}Al_{18}C_{36}$ | 108 | -125671.662 |
| $Cr_2TiAlC_2$ | $Cr_{Ti}$ | $Cr_{37}Ti_{17}Al_{18}C_{36}$ | 108 | -126537.2471 |
| $Cr_2TiAlC_2$ | $Ti_{Cr}$ | $Cr_{35}Ti_{19}Al_{18}C_{36}$ | 108 | -124805.2685 |
| $Cr_2TiAlC_2$ | $\{Cr_{Ti}:Ti_{Cr}\}_{11}$ | $Cr_{36}Ti_{18}Al_{18}C_{36}$ | 108 | -125670.847 |
| $Cr_2TiAlC_2$ | $\{Cr_{Ti}:Ti_{Cr}\}_{12}$ | $Cr_{36}Ti_{18}Al_{18}C_{36}$ | 108 | -125670.933 |
| $Cr_2TiAlC_2$ | $\{Cr_{Ti}:Ti_{Cr}\}_{13}$ | $Cr_{36}Ti_{18}Al_{18}C_{36}$ | 108 | -125670.7243 |
| $Cr_2TiAlC_2$ | $\{Cr_{Ti}:Ti_{Cr}\}_{21}$ | $Cr_{36}Ti_{18}Al_{18}C_{36}$ | 108 | -125670.854 |
| $Cr_2TiAlC_2$ | $\{Cr_{Ti}:Ti_{Cr}\}_{22}$ | $Cr_{36}Ti_{18}Al_{18}C_{36}$ | 108 | -125670.8151 |
| $Cr_2TiAlC_2$ | $\{Cr_{Ti}:Ti_{Cr}\}_{23}$ | $Cr_{36}Ti_{18}Al_{18}C_{36}$ | 108 | -125670.8765 |
| $Ti_3AlC_2$ | – | $Ti_{54}Al_{18}C_{36}$ | 108 | -94484.791 |
| $Ti_3AlC_2$ | – | $Ti_{96}Al_{32}C_{64}$ | 192 | -167972.791 |
| $Ti_3AlC_2$ | $Cr_{Ti}$ *(2a)* | $Ti_{53}CrAl_{18}C_{36}$ | 108 | -95350.36503 |
| $Ti_3AlC_2$ | $Cr_{Ti}$ *(4f)* | $Ti_{53}CrAl_{18}C_{36}$ | 108 | -95350.99609 |
| $Ti_3AlC_2$ | $Cr_{Ti}$ *(4f)* | $Ti_{95}CrAl_{32}C_{64}$ | 192 | -168839.0107 |
| $Ti_3AlC_2$ | $2Cr_{Ti}$ *(2a-2a)* | $Ti_{52}Cr_2Al_{18}C_{36}$ | 108 | -96217.14738 |
| $Ti_3AlC_2$ | $2Cr_{Ti}$ *(2a-2a)* | $Ti_{52}Cr_2Al_{18}C_{36}$ | 108 | -96217.1962 |
| $Ti_3AlC_2$ | $2Cr_{Ti}$ *(2a-2a)* | $Ti_{52}Cr_2Al_{18}C_{36}$ | 108 | -96217.13591 |
| $Ti_3AlC_2$ | $2Cr_{Ti}$ *(4f-2a)* | $Ti_{52}Cr_2Al_{18}C_{36}$ | 108 | -96216.48761 |
| $Ti_3AlC_2$ | $2Cr_{Ti}$ *(4f-2a)* | $Ti_{52}Cr_2Al_{18}C_{36}$ | 108 | -96216.41767 |
| $Ti_3AlC_2$ | $2Cr_{Ti}$ *(4f-2a)* | $Ti_{52}Cr_2Al_{18}C_{36}$ | 108 | -96216.49898 |
| $Ti_3AlC_2$ | $2Cr_{Ti}$ *(4f-2a)* | $Ti_{52}Cr_2Al_{18}C_{36}$ | 108 | -96216.54182 |
| $Ti_3AlC_2$ | $2Cr_{Ti}$ *(4f-2a)* | $Ti_{52}Cr_2Al_{18}C_{36}$ | 108 | -96216.54046 |
| $Ti_3AlC_2$ | $2Cr_{Ti}$ *(2a-2a)* | $Ti_{94}Cr_2Al_{32}C_{64}$ | 192 | -169705.2019 |
| $Ti_3AlC_2$ | $2Cr_{Ti}$ *(2a-2a)* | $Ti_{94}Cr_2Al_{32}C_{64}$ | 192 | -169705.2582 |
| $Ti_3AlC_2$ | $2Cr_{Ti}$ *(2a-2a)* | $Ti_{94}Cr_2Al_{32}C_{64}$ | 192 | -169705.1664 |
| $Ti_3AlC_2$ | $2Cr_{Ti}$ *(2a-2a)* | $Ti_{94}Cr_2Al_{32}C_{64}$ | 192 | -169705.2091 |
| $Ti_3AlC_2$ | $2Cr_{Ti}$ *(4f-2a)* | $Ti_{94}Cr_2Al_{32}C_{64}$ | 192 | -169704.617 |
| $Ti_3AlC_2$ | $2Cr_{Ti}$ *(4f-2a)* | $Ti_{94}Cr_2Al_{32}C_{64}$ | 192 | -169704.543 |
| $Ti_3AlC_2$ | $2Cr_{Ti}$ *(4f-2a)* | $Ti_{94}Cr_2Al_{32}C_{64}$ | 192 | -169704.623 |